\documentclass[a4paper]{jpconf}
\usepackage{graphicx}

\begin{document}

\title{Modeling of radiative and quantum electrodynamics effects in PIC simulations of ultra-relativistic laser-plasma interaction}

\author{M Lobet$^{1,2}$, E d'Humi\`eres$^2$, M Grech$^3$, C Ruyer$^1$, X Davoine$^1$, L Gremillet$^1$}
\address{$^1$CEA, DAM, DIF, F-91297, Arpajon, France}
\address{$^2$Univ. Bordeaux-CNRS-CEA, CELIA, UMR 5107, 33405, Talence, France}
\address{$^3$LULI, \'Ecole Polytechnique-CNRS-CEA-Universit\'e Paris VI, UMR 7605, \'Ecole Polytechnique, 91128 Palaiseau, France
}

\ead{mathieu.lobet@cea.fr}

\begin{abstract}
Next generation of ultra-intense laser facilities will lead to novel physical conditions ruled by collective and quantum electrodynamics effects, such as synchrotron-like emission of high-energy photons and $e^+e^-$ pair generation. In view of the future experiments performed in this regime, the latter processes have been implemented into the particle-in-cell code \textsc{calder}.
\end{abstract}

\section{Introduction}


The forthcoming ten petawatt laser systems, VULCAN-10PW in the United Kingdom (300 J, 30 fs) and APOLLON in France (150 J, 15 fs), will pave the way to the exawatt ELI project aiming to reach intensities in excess of $10^{25}\ \mathrm{Wcm}^{-2}$.
This will lead to unexplored laser-matter scenarii, characterized by coupled plasma and quantum electrodynamics (QED) processes. In order to enhance these novel effects, a favourable configuration relies upon the interaction of the laser pulse with counter-propagating, multi-GeV electrons. At laser intensities up to $10^{23}\ \mathrm{Wcm}^{-2}$, the electron dynamics enters the so-called radiation-dominated regime characterized by strong production of high-energy photons via nonlinear Compton scattering. Interacting back with the laser field, these photons can lead to efficient production of $e^-e^+$ pairs (via the Breit-Wheeler mechanism). The resulting dense pair plasma, beyond its fundamental interest, can potentially modify the overall laser-plasma interaction \cite{QED}.

\section{Modeling}


The importance of QED effects on a relativistic electron of mass $m_e$, charge $e$ and energy $\varepsilon_e = \gamma _e m_e c^2$, and on a photon of energy $\varepsilon_\gamma = \gamma _\gamma m_ec^2$, both interacting with a monochromatic plane wave $(\mathbf{E},\mathbf{B})$, depends on the respective Lorentz-invariant quantum parameters:
\begin{eqnarray}
\chi_{e} = \frac{\gamma _e}{E_s} \left[ \mathbf{E_{\parallel}}^2/\gamma _e^2 + \left(\mathbf{E_{\perp}} + \mathbf{v_e}\times \mathbf{B}\right)^2 \right]^{1/2}, & & \chi_{\gamma} = \frac{\gamma _\gamma}{E_s} \left[ \left(\mathbf{E_{\perp}} + \mathbf{c}\times \mathbf{B}\right)^2 \right]^{1/2},
\label{eq:etachi}
\end{eqnarray}
where $E_s = m_e^2 c^3 / e \hbar \simeq 1.3 \times 10^{18}\ \mathrm{V/m}$ is the Schwinger field beyond which the electromagnetic field can no longer be treated classically.
The energy distribution of the production rate of photons is given by \cite{Theory}
\begin{eqnarray}
 \displaystyle{\frac{d^2N _\gamma}{dt d\gamma_{\gamma}}} &=& \frac{1}{\pi \sqrt{3}} \frac{\alpha_f}{\tau_c \gamma _e ^2} \left[ \int_{2y}^{+\infty}{K_{1/3}(s)ds} - \left(2 + 3\chi_\gamma y \right)K_{2/3}(2y) \right],
\label{eq:dNsync_dtdgammae}
\end{eqnarray}
where $\alpha_f \simeq 1/137$ is the fine structure constant, $\tau_c = \hbar /m_e c^2$ the Compton length and $y=\chi _ \gamma / \left[ 3 \chi _e (\chi _e - \chi _\gamma) \right]$.
For $\chi_e \ll 1$ ($I < 10^{22}\ \mathrm{Wcm}^{-2}$, $\gamma _e \sim 2000$), large numbers of photons are emitted, each taking away a small fraction of the electron energy. The emission process can be therefore treated classically and modelled as a continuous damping reaction. 
By constrast, in the quantum regime, for $\chi_e \sim 1$ ($I \sim 10^{23}\ \mathrm{Wcm}^{-2}$), fewer photons are emitted, but their energies can be comparable to that of the electron, leading to strong recoils and discontinuous trajectories.
Finally, for high-energy photons ($\varepsilon_\gamma > 2 m_e c^2$) interacting with the laser field, the energy distribution of the probability to decay into $e^+e^-$ pair is given by \cite{Theory}
\begin{eqnarray}
\frac{d^2N_{\pm}}{dt d \gamma_e } = \displaystyle{ \frac{1}{\pi \sqrt{3}} \frac{\alpha _f }{ \tau _c \gamma _{\gamma}^2 } \left[ \int_{2y}^{+\infty}{ K_{1/3}\left( s \right)} ds - \left( 2 - 3 \chi_{\gamma} y \right) K_{2/3} \left(2 y \right) \right] } .
\label{eq:dNpm_dtdgamma}
\end{eqnarray}
Here, $\gamma_e$ refers to the energy of the created electron.

\section{Numerical implementation}

Particle-in-cell (PIC) codes are efficient tools for simulating the kinetic evolution of plasmas subjected to intense electromagnetic waves \cite{PIC}. The PIC code \textsc{calder} has been enriched with physical models describing photon and pair generation. In the PIC description, particles are represented by super-particles with given density, shape and momenta. Radiative and QED models are applied on the super-particles and we similarly define the notion of super-photons generated during the emission process.
In the classical radiative regime, the model considered is based on Sokolov's scheme for radiation damping \cite{Continuous}:
\begin{eqnarray}
\frac{d \mathbf{p_e}}{dt}  =  \mathbf{f}_l - e \left( \delta \mathbf{v_e} \times \mathbf{B} \right) - P_{rad} \frac{ \mathbf{p_e} }{\gamma m_e c^2}, & \displaystyle{ \frac{d \mathbf{x_e}}{dt} } =  \mathbf{v_e} + \delta \mathbf{v_e}, & \delta \mathbf{v_e}  = \frac{\tau _0}{m_e} \frac{ \mathbf{f}_{l} - \mathbf{v_e} \left( \mathbf{v_e} \cdot \mathbf{E} \right) }{1 + \tau_0 ( \mathbf{v_e} \cdot \mathbf{E} ) / m_ec^2}
\label{eq:sokolov_eq_three_vector}
\end{eqnarray}
where $\mathbf{f}_l$ is the Lorentz force vector and $\tau_0 = 2e^2/(12\pi \varepsilon_0m_ec^3)$.
The radiated power is calculated using the quantum equation (\ref{eq:dNsync_dtdgammae}):
\begin{eqnarray}
P_{rad} = \int_{1}^{\gamma_e} \frac{d^2 N_\gamma}{dt d\gamma} \gamma m_e c^2d\gamma
\end{eqnarray}

In the quantum radiative regime, the Monte Carlo (MC) description is well-suited to treating the stochastic character of the photon generation \cite{Discontinuous}. This is valid provided that the field is uniform and quasi-static during the emission process. Propagating electrons are assigned an optical depth $\tau_e$, evolving with time according to the field and the electron energy,
\begin{eqnarray}
\frac{d \tau _e}{dt} = \int_{0}^{\chi _e}{ \frac{d^2N}{d\chi dt} d\chi}.
\label{eq:tau_e}
\end{eqnarray}
In the MC algorithm, a discrete photon is emitted when $\tau_e = \tau_e^f$, where the total optical depth $\tau_e^f$ is sampled from $\tau_e^f = -\log{\xi}$ with $\xi \in ] 0,1]$ a uniform random number. The photon quantum paramater $\chi_\gamma$ is calculated by inverting $P_\gamma(0 \rightarrow \chi _\gamma) = \xi'$ where 
\begin{eqnarray}
P_\gamma(0 \rightarrow \chi ) = { \displaystyle{ \int_{0}^{\chi}{ \frac{d^2N_\gamma}{dtd\chi}  d\chi }} }/{ \displaystyle{\int_{0}^{\chi _e}{ \frac{d^2N_\gamma}{dtd\chi} d\chi}}}
\label{eq:P}
\end{eqnarray}
is the the photon cumulative distribution function and $\xi' \in \left[0,1\right]$ is a new random parameter.
The photon is emitted to the direction parallel to the electron velocity with wave vector $\mathbf{k}$ (since the angular spread is $\sim 1/\gamma_\gamma \ll 1$). The electron momentum $p_e^f$  therefore becomes
\begin{eqnarray}
\mathbf{p}^{f}_e = \mathbf{p}_e - \hbar \mathbf{k} = \displaystyle{ \left( p_e - \frac{\varepsilon _{\gamma}}{c} \right) \frac{\mathbf{p}_e}{\left| \mathbf{p}_e \right|}}.
\label{eq:pf}
\end{eqnarray}
A similar MC scheme is applied to the super-photons to model pair creation using the probability function (\ref{eq:dNpm_dtdgamma}). Each photon is assigned an optical depth $\tau_\gamma$ and a randomly sampled final one $\tau_\gamma^{f}$.When $\tau_\gamma = \tau_\gamma^f$, the super-photon is deleted and a $e^-e^+$ pair is created at the current position. Solving  $P_{\pm}(0 \rightarrow \chi _-) = \xi''$ gives the newly-created electron quantum parameter $\chi_-$, $\xi''$ being randomly calculated in $\left]0,1\right]$ and $P_{\pm}$ the cumulative distribution function associated to the pair generation. The positron quantum parameter is $\chi_{+} = \chi_{\gamma} - \chi_{-}$.
As the MC radiative model leads to the generation of many low-energy photons unable to decay into pairs and prejudicial for the numerical efficiency (most notably in configurations prone to pair cascading), we usually restrict the creation of super-photons to energies above $2m_ec^2$. Moreover, we have worked out a coalescence scheme ensuring energy and momentum conservation on average.

\section{Validation test}

As a test simulation, we consider the photon and pair generation by electrons of initial energy $\varepsilon_e/m_ec^2 = 3000$ in a constant magnetic field $B=7.48\ \mathrm{MT}$. The initial quantum parameter is equal to $\chi _e \simeq 5$. The initial synchrotron frequency and radius are respectively equal to $\omega_B^{-1} = 2.3\ \mathrm{fs} $ and $R \simeq 0.7\ \mu \mathrm{m}$. We define $f_{+}, f_-, f_\gamma$ the energy distributions of the positrons, electrons and photons, respectively. The time evolution of the latter can be described by the following set of equations \cite{Distribution}
\begin{eqnarray}
\frac{d f_{\gamma}}{dt} (\gamma _{\gamma},t) &=& \int_{\gamma_\gamma}^{+\infty}{ \frac{d^2N_{\gamma}}{dt d \gamma _{\gamma} } (\gamma _{\gamma},\gamma) f_\pm (\gamma) d \gamma } - 2\int_{2}^{\gamma_{\gamma}}{\frac{d^2N_{\pm}}{dt d \gamma _{\pm} } (\gamma ,\gamma _{\gamma}) f_{\gamma} (\gamma_{\gamma}) d \gamma }  \label{eq:spc_th1} \\
\frac{d f_{\pm}}{dt } (\gamma _{\pm},t) & = & -\int_{0}^{\gamma _{\pm}}{ \frac{d^2N_{\gamma}}{dt d \gamma _{\gamma} } (\gamma _{\pm},\gamma) f_{\pm} (\gamma _\pm) d \gamma } + \int_{\gamma_{\pm}}^{+\infty}{ \frac{d^2N_{\gamma}}{dt d \gamma _{\gamma} } (\gamma ,\gamma - \gamma _{\pm})  f_{\pm} (\gamma) d \gamma }  \nonumber \\ 
 & & + \int_{\gamma _\pm+1}^{+\infty}{ \frac{d^2 N_{\pm}}{ dt d \gamma _{\pm} } (\gamma _{\pm}, \gamma) f_ \gamma (\gamma) d\gamma } \label{eq:spc_th2}. 
\end{eqnarray}
\begin{figure}[t]
\begin{center}
\includegraphics[width=0.85\textwidth]{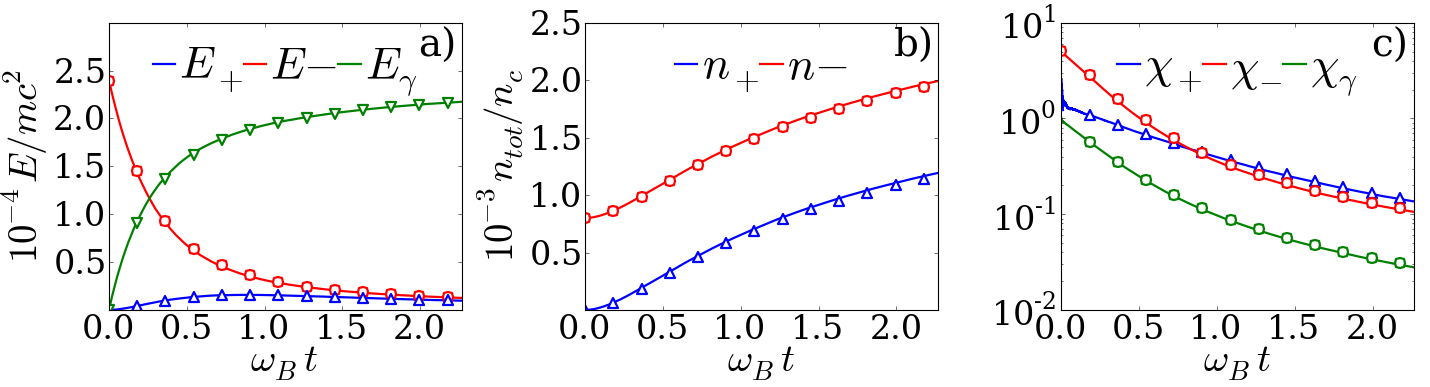}
\caption{Time evolution of the total energy (a), density (b) and quantum parameter (c) from the PIC simulation (lines) and the numerical solution of Eqs. (\ref{eq:spc_th1}) and (\ref{eq:spc_th2})  (markers).}
\label{fig:Time_diag_group}
\includegraphics[width=0.87\textwidth]{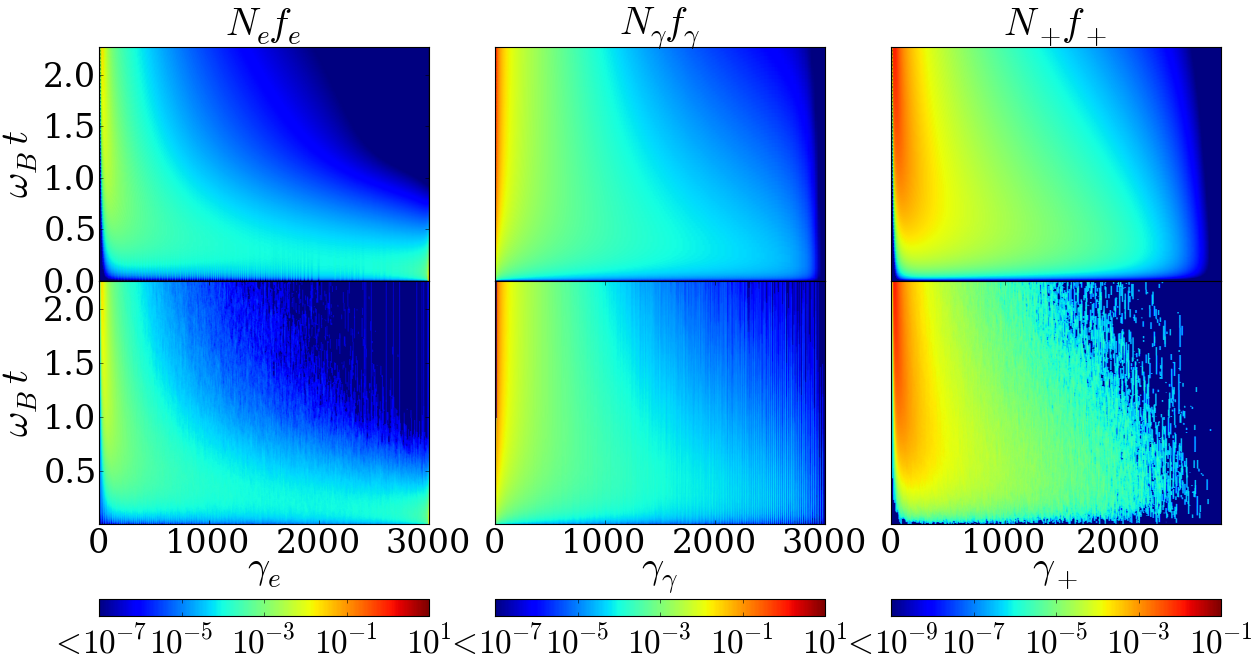}
\caption{Time evolution of the energy distributions for electrons (right), photons (center) and positrons (left) from the PIC simulation (bottom) and the numerical solution of Eqs. (\ref{eq:spc_th1}) and (\ref{eq:spc_th2}) (top).}
\label{fig:Spc_group}
\end{center}
\end{figure}
The time evolution of the total energy, density and averaged quantum parameter is plotted in Fig. \ref{fig:Time_diag_group} for each species. Figure \ref{fig:Spc_group} displays the corresponding energy distributions. Very good agreement is obtained between the numerical solutions of Eqs. (\ref{eq:spc_th1}) and (\ref{eq:spc_th2}) and the MC-PIC simulations.
The interaction starts in the quantum regime. The firstly created photons have an energy close to initial electron energy, $\gamma _{\gamma} \sim 10^3$. Consequently, the radiated energy significantly grows during the first rotation at the expense of the electron energy (Fig. \ref{fig:Time_diag_group}a). The average photon quantum parameter is close to $\chi_{\gamma} \sim 1$ (Fig \ref{fig:Time_diag_group}c), which is favourable for pair production. In turn, newly generated particles radiate their energy away, hence leading to pair cascading (Fig. \ref{fig:Time_diag_group}b).
When entering the semi-classical regime, the production rate of positrons tends to stabilize since the average photon energy drops below the pair creation threshold ($\chi_{\gamma} < 10^{-1}$). Radiation losses, originally responsible for pair creation, now only act as a cooling mechanism.
Note that in the quantum regime, the stochasticity, intrinsic to the QED processes stretches the electron energy spectrum, whereas in the semi-classical regime, radiation losses tend to shrink it back.

\section{Conclusion}
We have presented the implementation in a PIC code of the photon and pair generation via the Compton scattering and Breit Wheeler processes. These physical models will be used in forthcoming papers to study a number of laser-plasma scenarii at ultra-relativistic intensities.

\ack
The authors acknowledge support by the French Agence Nationale de la Recherche (LABEX PALM-ANR-10-LABX-39) and thanks R Duclous, A Debayle for interesting discussions. The PIC simulations were performed using HPC resources at TGCC/CCRT (Grant No. 2013-052707).

\end{document}